# ‘A BIT OF CHAOS AND MADNESS’: THE AI ASSESSMENT SCALE AND THE WORK OF ASSESSMENT REFORM

A PREPRINT

Mike Perkins [1*], Darius Postma [1], Jasper Roe [2], Susan Sisay [3], Craig Holdcroft [3]

[1] British University Vietnam, Vietnam
[2] Durham University, United Kingdom
[3] University of Staffordshire, United Kingdom

[*] Corresponding Author: mike.p@buv.edu.vn

June 2026

## Abstract

Generative artificial intelligence (GenAI) has intensified pressure on universities to redesign assessment while maintaining integrity, equity, and validity. Structured frameworks such as the Artificial Intelligence Assessment Scale (AIAS) offer one response, but evidence of how staff experience their implementation remains limited. This qualitative study examines AIAS implementation at a private international university in Vietnam and a public university in the United Kingdom. Data from five focus groups with 30 academic staff were analysed using hybrid thematic analysis, with Critical AI Literacy used as a sensitising concept. Six themes were developed: recognising and integrating AI, facilitating conditions, building capacity, pathways to adoption, ethics in practice, and reframing pedagogy.

Staff valued the AIAS as a shared language for legitimising GenAI use, clarifying boundaries, and prompting reflection on assessment design. However, implementation was shaped by governance, tool access, staff confidence, workload, integrity concerns, disciplinary context, and alignment with learning outcomes. The findings show that the AIAS could prompt authentic assessment design and student engagement, but may become a compliance layer when disconnected from learning outcomes, disciplinary context, and staff capacity. This study contributes empirical evidence on the institutional conditions through which GenAI assessment frameworks move from policy adoption to pedagogical enactment.

# Introduction

Artificial Intelligence (AI) represents the latest phase of digital automation in higher education, which has repeatedly reshaped pedagogical labour and governance (Selwyn et al., 2023). In terms of assessment, regulatory and sector guidance has increasingly framed this as a reform challenge, arguing that integrity, validity, and transparency must be built into assessment design and governance rather than retrofitted through enforcement (Lodge et al., 2023; Lodge, 2024). Early institutional responses to generative AI (GenAI) tools tended to focus on prohibition and detection (An et al., 2025); however, evidence shows that these approaches are limited, inequitable, and easily bypassed (Sun et al., 2026; Roe, Perkins, Bannister, et al., 2026). Ardito (2025) similarly argues that detection is misaligned with the trajectory of higher education, where students must learn to engage responsibly with GenAI rather than avoid it. This critique reflects a broader repositioning within the field regarding whether assessments still support valid claims about what students know and can do (Dawson et al., 2024), as opposed to the artefacts they can produce using GenAI tools.

Policy framing also requires scrutiny. Luo (2024) shows that many institutional policies construct GenAI as a threat to ‘originality’, reinforcing outdated notions of authorship while ignoring the collaborative and mediated nature of knowledge production. Perkins and Roe (2025) highlight how GenAI risks amplifying inequalities of access and reshaping epistemic norms, while Corbin, Dawson, and Liu (2025) caution that institutional responses often remain rhetorical, consisting of policies, declarations, and disclaimers, without meaningful reform of assessment tasks. They argue that genuine change demands redesign rather than discursive fixes. At the sector level, a consensus is beginning to emerge. The Russell Group (2023) has articulated principles emphasising AI literacy, equity of access, authentic assessment and staff capacity. These signal a shift beyond prohibition toward integration and shared responsibility but remain aspirational until enacted. In practice, universities have operationalised this shift through a range of responses, including increasing the use of secured assessment components for assurance and adopting framework-based approaches to standardise guidance on the use of GenAI. Common approaches include scale-based frameworks, such as the AI Assessment Scale (AIAS) (Perkins, Furze, et al., 2024; Perkins et al., 2025b), lane-based models, such as the two-lane approach (Liu & Bridgeman, 2023), and traffic light-style categorisations of ‘allowed’ AI use. Wherever such lines are drawn, however, they must still be interpreted by staff and students, who can experience the boundaries of acceptable use as inconsistent or arbitrary (Corbin, Dawson, Nicola-Richmond, et al., 2025). The risks these initiatives address are not only institutional but also epistemic. As Larson et al. (2024) argue, GenAI threatens both the individual and social dimensions of critical thinking, highlighting the need for assessments that cultivate scepticism and epistemic reflection rather than uncritical reliance.

The AIAS, developed by Perkins, Furze, et al. (2024), provides structured guidance for embedding GenAI into assessment design across the dimensions of integrity, equity, and pedagogy. The framework has since been revised to clarify its theoretical basis and strengthen its function as a validity-oriented redesign tool (Perkins et al., 2025b); however, empirical evidence of how faculty experience its implementation across institutional contexts remains limited. Ateşkan (2026) provides an initial empirical investigation examining how graduate educators at a single Turkish university conceptualised and applied the AIAS in AI-integrated lesson plan design. The study found that selecting the appropriate level of the AIAS functioned as an ethical boundary-setting practice, supporting educators in making judgements about delegating evaluative responsibility between human and AI agents. Although this is an important first step, there are substantial empirical gaps.

This study addresses this empirical gap by examining the practical implementation of the AIAS across two institutional contexts. To achieve this, we conducted five focus groups with 30 academic staff across two institutions to address the following research questions:

**RQ1:** How do faculty experience the implementation of the AIAS?

**RQ2:** Which specific factors enable or hinder the implementation of the AIAS and similar frameworks?

**RQ3:** What forms of support can help prepare staff for ongoing technological disruption?

As a result, this paper contributes to the assessment literature in three ways. First, it provides evidence of how institutional and faculty perspectives shape the enactment of assessment frameworks across institutional settings, moving beyond conceptual debates. Second, it situates these experiences within the wider sectoral discourse, arguing that sustainable assessment reform requires moving beyond detection and prohibition towards policies that balance integrity, equity, and innovation. Third, the study makes a conceptual contribution by shifting attention from the design of assessment frameworks to the conditions of their enactment.

## Methodology

Participants were drawn from two institutions that had previous experience at the faculty level with the AIAS: a private international university in Vietnam [I1], which had adopted the first version of the AIAS institution-wide, and a public university in the United Kingdom [I2], which had adopted the second version in one school. Focus groups were chosen as the primary method for data collection, and a semi-structured protocol was employed to balance consistency with flexibility for emergent discussions. Purposive sampling across three focus groups was used in I1 to ensure an appropriate spread of faculty experiences across disciplines (n = 17). In I2, self-selection was used for the two focus groups as participation was limited to the single school in which the AIAS had been trialled (the Business School) (n = 13). The research team co-developed the questions through iterative discussions to align with the study’s aims and institutional contexts (Creswell & Poth, 2018). Each session lasted approximately 90 minutes, was audio-recorded with consent, and was conducted in person at I1 and online at I2. The focus groups consisted of teaching-focused academic staff who had experience in operationalising the AIAS in their assessments, and the sessions were conducted by peers who were not involved in the development of the AIAS. Ethical approval was obtained from both institutions, and the data were anonymised prior to analysis.

The analysis followed a hybrid thematic approach, combining deductive coding with inductive theme development from focus group data. Our deductive coding framework was developed using the research questions of the study and the focus group protocol, with Critical AI Literacy (CAIL; Roe, Perkins, & Furze, 2025) used as a sensitising concept to orient the codes relating to ethics, equity, and criticality. Categories were agreed upon during protocol development and refined during early coding, following principles of hybrid thematic analysis (Fereday & Muir-Cochrane, 2006). These codes provided an initial analytical structure, while the final themes were developed inductively through the iterative interpretation of patterns of meaning across the transcripts from both institutions, consistent with Braun and Clarke’s (2022) emphasis on themes as analytic constructions rather than simple summaries of coded material. The full deductive coding framework is presented in Appendix A.

**Table 1: Deductive Coding Framework: Parent Codes**

| Code ID | Parent Code | Description |
|---|---|---|
| PC1 | Infrastructure and Resources | Comments about availability, accessibility, or constraints related to AI tools, subscriptions, and technical infrastructure. |
| PC2 | Policy and Implementation | References to institutional strategies, policies, rollout processes, and clarity of communication. |
| PC3 | Faculty Preparedness and Confidence | Faculty perceptions of readiness, confidence, and attitudes. |
| PC4 | Barriers and Challenges | Obstacles or constraints preventing effective use. |
| PC5 | Assessment Design and Alignment | Perceptions of how the AIAS aligns with learning outcomes and assessment. |
| PC6 | Impact on Learning and Engagement | Perceived effects on student behaviour and learning. |
| PC7 | Training and Support Needs | Perceived gaps or needs in training and professional development. |
| PC8 | Ethical and Equity Considerations | Fairness, cheating, and ethical challenges. |
| PC9 | Critical AI Literacy | Ability to critically evaluate AI outputs and understand limitations. |
| PC10 | Recommendations and Future Directions | Suggestions for improving the AIAS, training, or policy. |

Primary coding was performed by one member of the research team. This approach was appropriate for the study’s hybrid thematic design, where the purpose was to generate an interpretive account of the lived experience of academic staff rather than to establish inter-coder reliability. To strengthen analytic transparency and reduce the risk of institutional bias, an external co-researcher independent of both institutions reviewed the coding framework and its application to a subset of coded extracts from both datasets. The research team discussed points of divergence and used them to refine the code definitions, clarify the boundaries between the codes, and test the coherence of the developing themes. Reflexivity was maintained through memoing, journaling, and team discussions to interrogate assumptions and mitigate bias, consistent with Finlay’s (2002) dialogical reflexivity. In relation to positionality, two members of the author team were among the developers of the AIAS, and I1 was the setting in which the AIAS had originally been developed. This afforded familiarity with the framework's intended use but carried a risk of confirmation bias. To mitigate this, focus groups and initial rounds of coding were facilitated by another co-researcher.

# Results

Through iterative synthesis, the initial 10 parent codes were collapsed into six overarching themes, as patterns of meaning were present across multiple code categories. Following Braun and Clarke (2022), they are treated as interpretive patterns that cut across codes, offering a more complete account of how faculty engaged with GenAI and the AIAS. The themes highlight both convergences and divergences between the private university in Vietnam and a public university in the UK. These differences are treated as alternative vantage points derived from the differing contexts. Each theme is composed of several sub-themes that capture the more specific patterns that comprise it, as shown in Table 2.

The six themes summarised in Table 2 capture the distinct but connected dimensions of how the faculty made sense of the integration of GenAI and AIAS into their assessment contexts.

**Table 2: Thematic Structure of Results**

| Theme | Sub-themes | Interpretive focus |
|---|---|---|
| 1. Recognising and Integrating AI | Curiosity and experimentation; Legitimacy; Uneven integration | The scale lent legitimacy to bringing GenAI into teaching and assessment; however, integration was uneven and often credited to GenAI in general rather than the scale itself. |
| 2. Facilitating Conditions | Negotiating ambiguity; Constraint | Policy gaps and unclear mechanisms created uncertainty; legitimacy was granted by the AIAS but confidence and fairness were undermined. |
| 3. Building Capacity | Training; Support needs | Confidence grew through practice-based training, peer exchange, and experimentation, seen as vital for developing critical awareness. |
| 4. Pathways to Adoption | Comparative models; Governance models | Governance shaped adoption: rapid rollout generated momentum but curtailed reflection, while departmental leadership filled gaps but fragmented strategy. |
| 5. Ethics in Practice | Ethics; Equity; Critical AI Literacy | Ethics was framed through fairness and critical literacy, with equity and questioning viewed as essential to responsible adoption. |
| 6. Reframing Pedagogy | Assessment redesign; Student accountability | The AIAS made GenAI use visible within assessment design, prompting staff to negotiate tensions between authentic engagement, student accountability, disciplinary fit, and alignment with learning outcomes. |

### Theme 1: Recognising and Integrating AI

Across both institutions, faculty engagement with GenAI was characterised by curiosity and experimentation, with faculty describing the AIAS as a framework that legitimised these emerging practices in higher education. One early career lecturer welcomed having *"an assessment scale that I can base on and not having to do it on my own… a scale there to help me navigate this"* (I1: FCG3 – SP3).

At I2, innovation was more structured and embedded. One lecturer explained, *"At the end of each session there was an activity where an AI tool is used… the impact on the student and on my teaching… it's quite… growing compared to the previous year"* (I2: FCG1 – SP3). In terms of supporting the use of GenAI tools, one participant said: "*The AI assessment scale was really useful because they had to declare… what they'd used and why they'd used it… it's really… interesting for me to see… how they were using it in terms of their learning*" (I2: FCG2 – SP6). Another described the confidence the scale gave: *"It's given the apprentices some reassurance that 'oh yeah, we are allowed to use this, and this is appropriate.' It's given me the same thing… it's given me some confidence that this is OK to use and it's fine for us to experiment with and we're not going to be reprimanded for… using something which is considered a cheat tool"* (I2: FCG2 – SP9).

In the early stages of adoption, engagement focused more on normalising GenAI use than critiquing it. Faculty described the AIAS as providing a form of bounded permission for GenAI use in assessment: it made some experimentation feel legitimate and moved practice beyond simple prohibition, while also requiring staff and students to negotiate limits, declarations, and appropriate levels of use. For some participants, these boundaries supported more critical engagement with GenAI. Others, however, attributed their changing practice to GenAI in general rather than to the scale; as one lecturer put it, *“I have changed a lot because of AI but not because of the scale… except the assessment writing” (I1: FCG3 – SP2).*

**Theme 2: Facilitating Conditions**

Institutional environments shaped how staff managed uncertainty around GenAI and the AIAS, particularly regarding policy clarity, workload, and the ability to monitor student use.

At I1, staff described how unclear processes limited confidence. One lecturer reflected, *“Just not knowing who’s using AI and how… it’s a bit of chaos and madness… we’re already really put under pressure for the marking. So, making that more… laborious is an issue”* (I1: FCG2 – SP3). Another emphasised the lack of ability to properly evaluate how students were using AI tools, asking, *“It’s about students who see that on an AI scale 3… and then you don’t have anything to do about it… even we have included it in the rubric… how can we know what they are doing with it?”* (I1: FCG2 – SP5). Similar concerns were raised by the faculty at I2. One participant explained, *“We still… don’t have clear indication in terms of how we can apply it (GenAI), which sometimes… restrict you as an individual in terms of finding a way to even talk about artificial intelligence and its use with students”* (I2: FCG2 – SP8).

Against this uncertainty, other accounts, particularly at I2, described the scale itself as a source of clarity and a shared reference point. One participant contrasted expectations with experience: “*If it’s been developed by academics, it’s probably far too complex… and that’s not the case at all… It’s worded nice and simple. There’s four points [sic] on the scale, not 104…*” (I2: FCG2 – SP9). Several linked this clarity to the requirement to declare AI use, describing the scale as “really useful because they had to declare… what they’d used and why” (I2: FCG2 – SP6), as bringing “*more clarity in terms of how and in what way [students] can use AI*” (I2: FCG2 – SP5), and as a way to “*open up that conversation with students… rather than just leaving it open*” (I2: FCG2 – SP3). At I1, this clarity emerged over time rather than at the rollout: “*That was a developing stage… after that we did get some clarity, but initially… it was quite vague*” (I1: FCG3 – SP4).

Faculty accounts across institutions pointed to a shared tension, in which the AIAS legitimised GenAI use in both classrooms and assessment tasks, but the lack of clear policy and evaluative mechanisms left expectations for both faculty and students uncertain. This ambiguity not only eroded staff confidence but also raised concerns about whether fairness in assessment could be maintained in the future.

**Theme 3: Building Capacity**

At I1, staff emphasised that training needed to be directly connected to teaching practice. One lecturer explained, *“So, in the future, I think it would be beneficial… more specifically, teaching and learning… we are all learning as we are going now… perhaps six months to a year from now we’ll be in a stronger position”* (I1: FCG3 – SP3). Another reflected on the need for hands-on peer exchange: *“…personally, I would have liked… some practical stuff so I can see what people are really doing with it… much more hands-on… that would have been a big plus”* (I1: FCG2 – SP2).

At I2, participants also highlighted the importance of collegial support and practice-based learning in their development. One lecturer suggested, *“…perhaps sharing of best practice amongst the colleagues will be helpful as a form of guidance… if there’s some element of sharing that knowledge or best practice within the group… that will be really helpful”* (I2: FCG2 – SP5).

Effective capacity building was therefore strongest when faculty members had opportunities to engage with peers and share practices rather than relying on top-down directives and formal written communications. These conditions were seen as essential for building critical awareness of what might be possible with GenAI tools, and therefore supporting responsible adoption in assessment practices.

**Theme 4: Pathways to Adoption**

Faculty across both institutions described how the adoption of the AIAS reflected different governance models, with contrasting implications for confidence and consistency.

At I1, adoption was experienced as rapid and sometimes improvised. One lecturer reflected, *“So, I think because when we [I1] released the scale it was off the back of the release of ChatGPT… my feeling at the time was that the assessment scale was just released a little too early”* (I1: FCG1 – SP4). Another described how this timing left faculty struggling to keep pace: *“I think the problem is that everyone is dealing with this at the same time… it’s a race that’s ongoing and we can’t get ahead of it… we are chasing our own tails because it’s just developing at a rapid speed… in the future… perhaps six months to a year from now we’ll be in a stronger position”* (I1: FCG3 – SP6). Adoption was also experienced as a top-down expectation rather than a shared endeavour, with staff describing the scale as something encouraged by management largely to demonstrate that the institution had acted: *“Officially… we have some… encouragement by … the management, that yes, we can apply AI tools in our teaching and learning… But personally I have some questions about the effectiveness of that scale because we don’t have any mechanism to measure it… Except that that’s just to protect the interest of the university that at least we did something”* (I1: FCG3 – SP2).

At I2, governance was more decentralised and department-led, but still incomplete. One participant reflected, “*it’s decentralised and the Business School is doing things its own way… that’s a good thing… an opportunity for us to all explore this experiment…*” (I2: FCG2 – SP9). Others highlighted departmental leadership in filling gaps: *“It was more an internal departmental decision to adopt… the university didn’t have a policy… so we kind of led with this”* (I2: FCG2 – SP1).

Governance shaped adoption in distinct ways: a rapid rollout at I1 generated momentum but curtailed reflection and left limited time for faculty to understand the problem, while departmental leadership at I2 addressed policy gaps but left the strategy fragmented. Therefore, the contrast was one of governance logic: a top-down institutional mandate at I1, where faculty had limited input and a correspondingly weaker sense of ownership over the change, set against a bottom-up, department-led adoption at I2, where staff exercised greater local ownership but without an institution-wide direction.

**Theme 5: Ethics in Practice**

Faculty reflected on questions of equity and critical engagement with GenAI, highlighting how fairness and ethical practice were central to embedding GenAI in their teaching.

At I1, participants raised concerns regarding equity and criticality. One lecturer explained, *“I don't think, I mean, we try to address critical thinking throughout all our classes… skills how to be a critical thinker should be throughout your degree. And I don't think AI generated content is, in that respect, any different from wherever else content might be”* (I1: FCG2 – SP2). Another stressed the ethical challenge of superficial engagement: *“I have asked students when they come for feedback, can you review this?… what have you written in this paragraph? They don't know. So that’s the critical challenge, what is their output [if] based on AI”* (I1: FCG2 – SP5).

At I2, staff framed ethics more explicitly around critical engagement with GenAI. One participant noted, *“…at level 7 [Postgraduate] my students put that… I can read that through*

*and… they won't get marked because it doesn't… meet the assessment requirement"* (I2: FCG2 – SP7*)*. Another reflected on fairness in practice, stressing the problem of superficial engagement and lack of criticality*: "…it's very obvious they haven't got a clue about what they've even submitted, let alone written… they're literally taking whatever is generated… it's not critical… it's not referenced correctly… and it's very obvious that it's AI generated"* (I2: FCG1 – SP6). Equity of access was a further concern, with staff noting that the disparities GenAI introduces are not new but are still concerning, since students able to pay for more capable tools gain an advantage*: "If you put it onto books, some students can afford to buy books, and some not… So, they have more information than others. Same with ChatGPT" (I1: FCG1 – SP7).* One participant warned that *"the student that has access to and has paid for the upgraded version of ChatGPT, definitely has an advantage over the person who is using the free version… you're making it a tiered education system"* (I1: FCG1 – SP3).

Overall, the ethics of students using GenAI tools in their assessments was framed as a challenge of fairness and critical engagement rather than something that faculty needed to directly police. Faculty across institutions stressed that without equity and critical questioning, AI adoption risked becoming superficial and unsustainable, whereas embedding these principles was seen as essential for responsible integration.

**Theme 6: Reframing Pedagogy**

Faculty at both institutions described the AIAS as a framework that made GenAI use pedagogically visible, but its influence on teaching and assessment was uneven. For some participants, the scale prompted a reconsideration of assessment design, student accountability, and authentic engagement with AI. For others, it risked shifting attention away from intended learning outcomes, disciplinary skills, and established assessment purposes.

At I1, staff reflected on how the scale influenced assessment design. One lecturer explained, *"Because we are not writing the assessment for the scale. We are using the scale… for the outcome of the assessment. So now the focus is going on creating for the scale, but that should be the other way around. That's what I think"* (I1: FCG2 – SP5). Another noted the adjustments required: *"We should be more focused on specifically designing the assignment… if we are going in the previous way… we will fail… we… really need to review our… approach to writing the assessment and setting up the expectation"* (I1: FCG3 – SP2). Another account demonstrated how the AIAS could also be experienced as misaligning assessment when the level was retrofitted to an existing task: *"I felt that the AI assessment scale drove our assessment procedures away from learning outcomes… the higher tiers of the scale assess how students use the scale… rather than what we said that they were being assessed for"* (I1: FCG1 – SP3). Together, these accounts point to a central validity risk: when the AIAS level is treated as the object of design rather than as a means of aligning GenAI use with learning outcomes, the assessment may no longer generate the evidence of learning it was intended to capture.

At I2, the AIAS was seen as encouraging innovation. One participant reflected, *"For me it's made me think about… authentic assessments that actually assess learning rather than the ability to find that… when you're in a job, you're going to be able to just pull something up on AI… but actually, I think there's quite a lot around having that critical awareness"* (I2: FCG2 – SP3). Another emphasised responsibility: *"Although they've created it using AI tools, they are responsible for it. So, if AI tool creates rubbish, they are responsible for that rubbish… we do talk about the ethics, the responsible use of it, but we do that as a natural part of teaching AI"* (I2: FCG2 – SP4).

The AIAS did not appear to reshape pedagogy uniformly across all institutions. Instead, it prompted staff to reflect on what assessment should evidence in a context where GenAI use was increasingly difficult to separate from learning and professional practices. At I1, discussions centred on assessment design, with participants questioning whether AIAS levels were aligned with learning outcomes and whether assessment tasks were redesigned around the scale itself. At I2, the participants focused less on the mechanics of the framework and more on how it supported authentic assessment, critical engagement with AI, and student responsibility for AI-supported work. Taken together, the theme shows that the pedagogical value of the AIAS depended on how it was interpreted and embedded: it could support more accountable and authentic assessment practice, but it could also become a compliance layer when disconnected from learning outcomes, the disciplinary context, and staff capacity.

# Discussion

Across both institutions, faculty accounts pointed to a consistent message that higher education assessment reform was not secured through policy statements or directives alone, but through the conditions that made those policies workable in practice. Although frameworks such as the AIAS were described by faculty as providing a shared language and communicating shared expectations between faculty and students, participants repeatedly returned to the same institutional determinants of enactment: the pace and clarity of policy rollout, the extent of capacity building and support, the design work required of faculty to protect validity, and the equity implications of uneven access to GenAI tools and uneven staff preparedness to respond to GenAI. These findings reinforce sector warnings that discursive responses, such as declarations of “No AI” or generic compliance language, do not resolve the underlying validity problem if task conditions and marking criteria remain unchanged (Corbin, Dawson, & Liu, 2025; Dawson et al., 2024; Luo, 2024). In the following, we discuss governance, capacity, academic integrity, and equity, considering for each how they enabled or hindered implementation (RQ2) and what forms of support they imply for staff (RQ3).

### Governance: speed and clarity

The first set of factors shaping AIAS implementation concerned governance, specifically the relationship between the speed and clarity of policy roll-out. Faculty described operating in environments where GenAI adoption by students outpaced institutional guidance, generating uncertainty about permissible assessment practices and inconsistent messaging to students. Where policy moved quickly, staff reported momentum but also described “chasing” the technology and struggling to translate high-level intent into assessable task designs. Where policy was slower or fragmented, staff described filling gaps through departmental or individual decisions, which supported local innovation but amplified inconsistency and confusion across units. In both cases, the primary challenge was whether the institutions provided a coherent pathway from policy intention to assessable practice. These dilemmas illustrate why GenAI-related assessment has been characterised as a wicked problem: one that resists definitive formulation, admits only better or worse responses rather than correct ones, and places substantial responsibility on institutional decision-makers (Corbin, Bearman, et al., 2026). This implies that embedding structured faculty consultation within governance cycles related to GenAI and assessment from the outset is likely to be more productive than issuing top-down directives for compliance and may reduce reliance on the reactive guidance that has characterised the most rushed rollouts in higher education and beyond.

In terms of the support this implies, our two cases suggest that direction and ownership are complementary rather than competing: the rapid, centrally mandated rollout of the AIAS at I1 meant that there was momentum but limited faculty ownership, whereas the department-led approach at I2 built ownership but lacked broader institutional guidance. Therefore, effective governance appears to depend on combining a clear policy direction from the centre with support and encouragement from disciplinary champions closer to the practice. Making use of cross-disciplinary working groups, piloting major changes to assessment frameworks before mandating them, and building feedback loops that allow guidance to be revised as tools and student practice evolve are all areas that could support policy coherence and faculty ownership.

**Capacity and critical engagement**

Addressing RQ3, faculty were consistent in that the most useful form of support was sustained, practice-based capacity building rather than one-off briefings. Because assessment redesign in a GenAI context often requires new forms of judgement, for example deciding which learning outcomes require controlled conditions, where AI-supported work can be validly assessed, and how students should evidence responsible tool use, this is a key area for development. This aligns with broader accounts of the hidden labour involved in teaching with AI, including the ongoing “repair work” required to interpret, verify, and respond to AI-shaped student submissions and staff workflows (Selwyn et al., 2025). A practical model to integrate this might be to formalise faculty-led communities of practice or discipline-level “champions,” tasked with developing exemplar assessments, running peer workshops, and translating institutional expectations into locally meaningful tasks. Such structures distribute the design burden and allow for shared experiences to be transmitted more effectively around the institution, rather than leaving individual staff to solve the same problems in isolation. This may also give students more consistent practices across courses, rather than leaving them to navigate conflicting implementations of policy by different faculty.

**Academic integrity and validity**

Further factors concerned academic integrity and the limits of detection. Faculty in our focus groups were more concerned with whether current institutional strategies created fair and defensible assessment environments than with whether misconduct existed. Participants’ scepticism about monitoring and enforcement challenges resonates with evidence that AI text detection performs poorly and can be circumvented, while also creating risks of false accusations and inequitable impact on student groups (Liang et al., 2023). In this sense, the findings support a shift in emphasis from ‘catching’ AI use to designing for valid inference about learning (Dawson et al., 2024).

Closely related, the cases reported under Theme 6 expose the validity risk at the centre of framework implementation. When an AIAS level is applied to an existing task without design changes, the task may begin to assess students’ use of the framework or tool rather than the intended learning outcomes. Differing participant responses made it clear that simply implementing the AIAS without an effective means to ensure assessment design changes is not sufficient, and whether assessments actually changed was very much up to the individual academic. One participant described carrying assessments over with little redesign, noting that *“AI was there, but we didn’t change much in our assessment… we are giving level 4… because we didn’t update the time or assessment”* (I1: FCG3 – SP2). Yet, at the same institution, others used the AIAS to redesign tasks, with one explaining that it *“changed the way that I give them their… assignment briefs… [so that] they have to be authentic researchers and insert… explicit examples of how they did that research”* (I1: FCG1 – SP5). This pattern, in which levels are read as permissions attached to an otherwise unchanged task rather than as specifications of the task type and the evidence it should produce, empirically confirms a failure mode that the framework’s developers had anticipated but not previously evidenced (Perkins et al., 2025a). Institutions should therefore support assessment teams in making real design changes to assessment practices, including deriving decisions about GenAI-supported tasks from learning outcomes, ensuring disciplinary standards are maintained, and requiring suitable evidence that learning outcomes are being met, including how students will demonstrate evaluative judgement and accountability for AI-supported work (Dawson et al., 2024; Walton et al., 2025).

In terms of support, faculty accounts suggest that integrity and innovation are not in opposition to each other. Where staff framed AI use through responsibility, critique, and standards of evidence, they described clearer expectations with their students and more sustainable teaching practices than in contexts dominated by surveillance. Operationally, this points to replacing the language of policing with design-focused requirements: specifying where controlled conditions are needed to assure learning, where AI-supported work is permissible, and how rubrics should align with what each task can validly demonstrate. Assessment twins (Roe, Perkins, & Giray, 2026), in which a GenAI-vulnerable task is deliberately paired with a less vulnerable companion task targeting the same learning outcomes, offer one example of such triangulation, but the broader principle is to gather complementary evidence through task design rather than rely on the unenforceable promise that ‘no AI’ can be policed in take-home conditions.

**Equity and consistency**

The final set of factors centred on equity and consistency as conditions for policy legitimacy. Faculty repeatedly raised concerns that uneven access to tools, staff knowledge, and application of rules would disadvantage students and undermine their trust. These concerns align with Luo’s (2024) critique that policy framings centred on misconduct and surveillance can undermine trust by positioning students as potentially untrustworthy, teachers as gatekeepers, and legitimate GenAI use as less authentic or valuable. If universities want coherent reform, they need programme-level consistency that is compatible with disciplinary variation, meaning shared minimum expectations alongside the room for local design choices. Sector-level statements increasingly emphasise these principles, including staff capability building, equity of access, and a move towards authentic assessment that reflects real-world practice (Russell Group, 2023). However, the present study shows how quickly these ambitions break down when institutions do not invest in enacting such frameworks or policies.

**Beyond framework choice**

Taken together, these factors (RQ2) and the forms of support they imply (RQ3) highlight the contribution of structured frameworks such as the AIAS. Their value lies in supporting faculty to consider assessment design and functioning as a shared language through which faculty, students, and institutions can negotiate expectations; however, this function can only be realised when frameworks are enacted through structural redesign of tasks rather than adopted discursively (Corbin, Dawson, & Liu, 2025). This highlights a core theoretical contribution of this study: assessment validity does not emerge simply because the ‘correct’ framework has been chosen. More importantly, securing valid inferences about student learning depends on how any such framework is selected, implemented, supported, and sustained across an institution. This extends sector-level principles by specifying the institutional conditions, governance, capacity, and equity under which such frameworks can move assessment reform from aspiration to practice. These four sets of factors are not unique to the AIAS; they would apply to any framework. The study’s framework-specific contribution is therefore diagnostic, showing how a scale can degrade into a compliance exercise when retrofitted, and how its levels are read as permissions instead of task specifications.

**Limitations**

There are several limitations to this study. Methodologically, focus groups can be shaped by group dynamics and social desirability, constraining disclosure (Sim, 1998). A single-coder approach, required for practical reasons, precluded inter-rater statistics, although reflexivity and external validation mitigated this (O’Connor & Joffe, 2020). Sampling differed by site (purposive vs. self-selection), limiting comparability and under-representing other disciplines. Accordingly, the two institutions are treated throughout as contrasting vantage points on AIAS enactment, and we make no causal or comparative claims between sites; cross-site differences are read as situated illustrations of how context shapes implementation. The sites also used different versions of the AIAS (the original at I1, the revised version at I2), a difference that cannot be cleanly separated from national context, governance, timing, and scope. We therefore cannot attribute the more positive accounts at I2 to the revised scale, and a controlled comparison of the two versions is left to future work. The delivery mode of the focus groups also varied (in-person vs. online), which may have influenced interaction. Finally, the single cycle of data collection limited insight into longitudinal change. The study provides evidence of faculty perceptions and accounts of the conditions shaping AIAS implementation rather than measured generalisable outcomes; it does not establish whether the framework improved assessment validity, student learning, or integrity, and our claims are bounded accordingly.

## Conclusion

GenAI presents a complex and shifting set of challenges for educators, institutions, and students. Frameworks such as the AIAS can help institutions respond to these challenges by providing a shared language for AI use, assessment expectations, and student accountability. However, this study shows that their value is mediated by the conditions of implementation. The findings suggest that structured frameworks are most useful when they are embedded within sustained faculty development, consultative governance, equitable access, and programme-level consistency. Where these conditions are present, frameworks such as the AIAS can support more explicit conversations about GenAI use, encourage accountable student engagement, and prompt reconsideration of authentic assessment design. Where these conditions are absent, the same framework may become a compliance layer, with staff required to assign levels without sufficient confidence, support, or clarity about how those levels relate to valid evidence of learning.

The validity of assessment in a GenAI context is not secured by selecting a particular scale, level, or lane but depends on how that choice is interpreted, resourced, embedded, and sustained. When it is issued without these supports, it risks becoming another discursive response that leaves the underlying validity problem untouched (Corbin, Dawson, & Liu, 2025; Perkins et al., 2025a). Therefore, institutions’ practical priority is to invest in implementation: treating faculty development, equitable tool access, programme-level guidance, and assessment redesign capacity as core infrastructure for GenAI-era assessment reform.

### Declaration of Generative AI and AI-assisted technologies in the writing process

GenAI tools were used for ideation and in some passages of draft text creation, which was then substantially revised, along with editing, formatting, and refinement during the production of the manuscript. The tools used were ChatGPT (5.4, 5.5) and Claude (Sonnet 4.6, Opus 4.7, 4.8), which were chosen for their ability to provide sophisticated feedback on textual outputs. These tools were selected and used supportively and not to replace core author responsibilities. The authors reviewed, edited, and take full responsibility for all outputs of the tools used.

# Appendix A: Deductive Coding Framework (Parent and Sub-Codes)

| Code | Parent code / sub-code | Description |
|---|---|---|
| **PC1** | **Infrastructure and Resources** | **Comments about availability, accessibility, or constraints related to AI tools, subscriptions, and technical infrastructure.** |
| PC1.1 | Tool accessibility (free vs paid versions) | Disparities in access to free vs premium AI tools or subscriptions. |
| PC1.2 | Lack of licenses / subscriptions | Absent institutional provision of AI tools or accounts. |
| **PC2** | **Policy and Implementation** | **References to institutional strategies, policies, rollout processes, and clarity of communication.** |
| PC2.1 | Speed of rollout | Comments on rapid or rushed implementation timelines. |
| PC2.2 | KPI / compliance pressure | Pressure to comply with targets, KPIs, or mandates. |
| PC2.3 | Lack of clear guidance | Unclear instructions, inconsistent advice, or confusion about expectations. |
| **PC3** | **Faculty Preparedness and Confidence** | **Faculty perceptions of readiness, confidence, and attitudes.** |
| PC3.1 | Negative emotions and frustration | Stress, anxiety, or frustration linked to the AIAS. |
| PC3.2 | Lack of confidence in critical AI literacy | Not feeling equipped to evaluate AI outputs. |
| PC3.3 | Resistance to change | Reluctance to adopt AI tools. |
| **PC4** | **Barriers and Challenges** | **Obstacles or constraints preventing effective use.** |
| PC4.1 | Time and workload pressures | Lack of time or competing priorities. |
| PC4.2 | Confusion between levels (2-3) | Uncertainty distinguishing AIAS levels. |
| PC4.3 | Over-reliance by students | Concerns about students depending too much on AI. |
| **PC5** | **Assessment Design and Alignment** | **Perceptions of how the AIAS aligns with learning outcomes and assessment.** |
| PC5.1 | Misalignment with learning outcomes | The AIAS not supporting intended learning. |
| PC5.2 | Preference for authentic assessment | Desire for more authentic or practical assessment tasks. |
| PC5.3 | Limited student engagement with guidance | Students ignore AI guidance or support. |
| **PC6** | **Impact on Learning and Engagement** | **Perceived effects on student behaviour and learning.** |
| PC6.1 | Lack of critical evaluation | Students do not question AI outputs. |
| PC6.2 | Superficial use of AI | Shallow or convenience-based AI use. |
| PC6.3 | Limited reflection and declaration | Lack of transparency in AI use. |
| **PC7** | **Training and Support Needs** | **Perceived gaps or needs in training and professional development.** |
| PC7.1 | Lack of structured training | Absence of organised training. |
| PC7.2 | Need for peer learning / examples | Wanting examples or peer support. |
| PC7.3 | Requests for microlearning / guides | Bite-sized or accessible learning materials. |
| **PC8** | **Ethical and Equity Considerations** | **Fairness, cheating, and ethical challenges.** |
| PC8.1 | Contract cheating / humanising AI | Concerns about plagiarism or third-party use. |
| PC8.2 | Inequity of access | Unfair access to tools or resources. |
| PC8.3 | Ethical ambiguity | Uncertainty about what is acceptable use. |
| **PC9** | **Critical AI Literacy** | **Ability to critically evaluate AI outputs and understand limitations.** |
| PC9.1 | Limited faculty expertise | Lacking expertise to judge AI content. |
| PC9.2 | Lack of student awareness | Students do not understand AI limits. |
| PC9.3 | Difficulty encouraging critical use | Getting students to question AI outputs. |
| **PC10** | **Recommendations and Future Directions** | **Suggestions for improving the AIAS, training, or policy.** |
| PC10.1 | Consistent policies and training | Clearer policies and training alignment. |
| PC10.2 | Access to paid tools | Provide institutional access to paid AI tools. |
| PC10.3 | Clearer level definitions | Clarify AIAS levels. |